\documentclass[twocolumn,floatfix,preprintnumbers,nofootinbib,superscriptaddress]{revtex4}

\usepackage{ulem}
\usepackage{bm}
\usepackage{times}
\usepackage{amssymb,amsbsy,amsmath,amsfonts}
\usepackage{graphicx}
\usepackage{enumerate}
\usepackage{float}
\usepackage{color}
\usepackage{morefloats}
\usepackage{rotating}
\usepackage{srcltx}
\usepackage{slashed}
\usepackage{multirow}
\usepackage{verbatim}
\usepackage{hyperref}
\usepackage{tabularx}
\usepackage{braket}
\usepackage{amsfonts}
\usepackage{mathtools}
\usepackage{color}
\usepackage{mathrsfs}
\usepackage{gensymb}
\usepackage{booktabs}  
\usepackage{threeparttable}

\usepackage{changes}
\usepackage{xcolor}
\usepackage[outdir=./]{epstopdf}

\newcommand\be{\begin{eqnarray}}
\newcommand\ee{\end{eqnarray}}

\begin{document}
\title{Spin polarization and quantum entanglement of baryon-antibaryon pairs produced in electron-positron annihilation}

\author{Cheng Chen}
\affiliation{State Key Laboratory of Heavy Ion Science and Technology, Institute of Modern Physics, Chinese Academy of Sciences, Lanzhou 730000, China}
\affiliation{School of Nuclear Sciences and Technology, University of Chinese Academy of Sciences, Beijing 101408, China}

\author{Ju-Jun Xie}~\email{xiejujun@impcas.ac.cn}
\affiliation{State Key Laboratory of Heavy Ion Science and Technology, Institute of Modern Physics, Chinese Academy of Sciences, Lanzhou 730000, China} \affiliation{School of Nuclear Sciences and Technology, University of Chinese Academy of Sciences, Beijing 101408, China} \affiliation{Southern Center for Nuclear-Science Theory (SCNT), Institute of Modern Physics, Chinese Academy of Sciences, Huizhou 516000, China}

\begin{abstract}

In this work, we systematically investigate the evolution of spin polarization and quantum entanglement in cascade decays of baryon-antibaryon pairs, which are produced in electron-positron annihilation. We derive a fully analytical spin density matrix explicitly expressed in terms of spin polarization observables, extend this formalism to multi-step cascade decay scenarios, and establish compact recursive relations to facilitate density matrix calculations for such processes. It is found that when maximal parity violation occurs during a decay, the resulting final-state particles are fully polarized and exist in a non-entangled state. Furthermore, we demonstrate that quantum entanglement amplification is a generic characteristic of charge-conjugate decays under $CP$ conservation when the initially produced baryon-antibaryon pair is polarized.

\end{abstract}

\maketitle

\section{Introduction}

In quantum mechanics, spin represents a fundamental degree of freedom, and precise measurements of particle spin yield deep physical insights into the underlying interactions. Spin polarization quantifies the extent to which a particle’s spin aligns along a particular axis. In the process of electron-positron annihilation into a baryon-antibaryon pair $e^+e^-\to B\bar{B}$ ($B$ stands for baryon and $\bar{B}$ is antibaryon), analyzing the spin polarization of the final baryon (or antibaryon) allows the extraction of the relative phase between the electric and magnetic form factors of the final baryon~\cite{Faldt:2013gka,Faldt:2016qee,Faldt:2017kgy}. In Refs.~\cite{Perotti:2018wxm,Zhang:2024rbl} the polarization observables for final-state baryon pairs with various inherent spins produced in electron-positron annihilation were derived. Beyond single-particle polarization, baryon-antibaryon pairs produced in $e^+e^-$ annihilation are fundamentally entangled. Quantum entanglement represents one of the most profound departures of quantum mechanics from classical physics. To quantify the degree of entanglement in a two-qubit mixed state, the concept of entanglement of formation was introduced in Ref.~\cite{Bennett:1996gf}, a measure that naturally remains valid for pure states. Subsequently, an elegant and feasible approach for computing this quantity was developed~\cite{Hill:1997pfa, Wootters:1997id}, leading to the introduction of \textit{concurrence}. Since concurrence exhibits a monotonically increasing relationship with the entanglement of formation, it has become a standard metric for quantifying entanglement. 

Quantum entanglement has been extensively studied in connection with Bell's inequality~\cite{Bell:1964kc}. Investigations on Bell's inequality can experimentally demonstrate the non-locality of quantum systems, which constitutes a significant characteristic of entangled systems~\cite{Brunner:2013est}. Among the various forms of Bell's inequality, the Clauser-Horne-Shimony-Holt (CHSH) inequality is the most widely used~\cite{Clauser:1969ny}, and its rigorous relationship with entanglement was established in Ref.~\cite{Verstraete:2001skx}. Experimentally, Bell's inequality and entanglement have long been studied in photonic and atomic systems~\cite{Horodecki:2009zz,Giustina:2015yza,Stevens:2015awv,Rosenfeld:2017rka,BIGBellTest:2018ebd}. 

In high-energy physics, quantum entanglement has been observed in top-quark pair ($t\bar{t}$) production~\cite{ATLAS:2023fsd,CMS:2024pts}. Driven by advancements in experimental techniques and unprecedented statistical precision, the BESIII Collaboration has conducted extensive measurements of spin polarization in $e^+e^-\to \Lambda\bar{\Lambda}$~\cite{BESIII:2018cnd,BESIII:2021cvv,BESIII:2022qax,BESIII:2023euh}, $\Sigma\bar{\Sigma}$~\cite{BESIII:2020fqg,BESIII:2023ynq,BESIII:2024nif,BESIII:2024dmr}, and $\Xi\bar{\Xi}$~\cite{BESIII:2021ypr,BESIII:2022lsz,BESIII:2023drj,BESIII:2023lkg,Liu:2023xhg} reactions. These rich experimental findings have stimulated further theoretical investigations~\cite{Salone:2022lpt,Cao:2024tvz,Lin:2025eci,Tang:2025oav,Zhang:2025oks,Chen:2025xci,Chen:2024luh,Wu:2025dds,Li:2026bkf} on the spin polarizations and the quantum entanglement of the final produced baryon-antibaryon pair. Recently, the BESIII collaboration confirmed a violation of Bell's inequality in the $\Lambda\bar{\Lambda}$ system from the $\eta_c\to \Lambda\bar{\Lambda}$ decay process~\cite{BESIII:2025vsr}. Furthermore, in Ref~\cite{Fabbrichesi:2024rec}, both entanglement and Bell's inequality violation in charmonium decays were determined by analyzing the helicity amplitudes experimentally. In the $e^+e^-\to Y\bar{Y}$ process, studies have shown that the maximum Bell's inequality violation and entanglement occur at the scattering angle near $\pi/2$~\cite{Wu:2024asu}. In entangled multipartite systems, the decay of unstable particles transfers quantum information to their final-state products, leading to dynamic changes in the system's polarization and entanglement. A fascinating phenomenon known as "entanglement amplification" was recently identified in Ref.~\cite{Aguilar-Saavedra:2024fig}, where the decay of one particle in a spin-entangled $t\bar{t}$ pair can spontaneously increase the overall degree of entanglement. This phenomenon, known as “entanglement amplification,” has attracted theoretical interest. A similar phenomenon has been observed in baryon pair decays~\cite{Feng:2025ryr}.

In this paper, we systematically investigate the evolution of polarization and entanglement in baryon-antibaryon pairs during cascade decays. Specifically, we focus on spin-1/2 baryon pairs produced via $e^+e^-\to B\bar{B}$, followed by the weak parity-violating decays $B \to B_1^\prime + P$ and $\bar{B} \to \bar{B}_2^\prime + P$, where $P$ denotes a spin-0 pseudoscalar meson. Because the emitted mesons are spinless, the complete spin information is transferred to and preserved within the final $B_1^\prime\bar{B}_2^\prime$ subsystem. We will derive an explicit density matrix expressed entirely in terms of physical polarization observables and provide a fully analytical framework to track the entanglement evolution from the initial production to the final decayed state.
 
The remainder of this paper is organized as follows: In Sec.~II, we present the details of the derivation of the helicity amplitudes for both processes and the corresponding spin density matrix. In Sec.~III, we analyze the evolution of polarization and entanglement, presenting a set of conditions that determine their preservation, enhancement, or suppression. Finally, a brief summary is provided in Sec.~IV.

\section{Theoretical Framework}

We start from the density matrix for the $B\bar{B}$ pair in the process $e^+e^-\to B\bar{B}$, together with the corresponding polarization observables. We then incorporate the cascade decays of $B$ and $\bar{B}$ to derive the final-state density matrix, and analyze the spin polarization and entanglement. Finally, we present recursive relations for the density matrix and angular distribution in multistep cascade decays of the final baryon $B$ and antibaryon $\bar{B}$.

\begin{figure}[htbp]    
    \centering
    \includegraphics[scale=0.55]{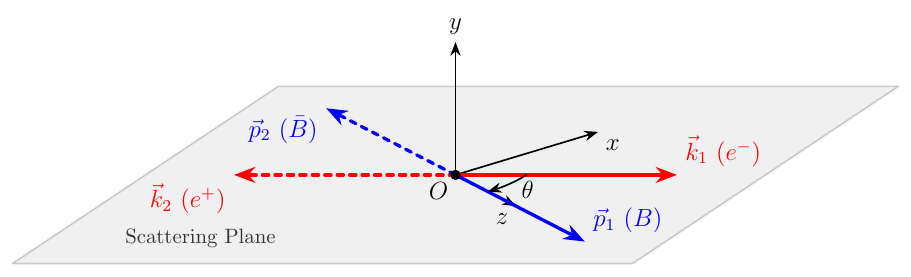}
    \caption{Definition of the coordinate system for the $e^+e^- \to B \bar{B}$ reaction in the center-of-mass frame. The $z$-axis is chosen along the momentum of the outgoing baryon $B$, and the $y$-axis is perpendicular to the scattering plane defined by $\vec{p}_1 \times \vec{k}_1$. The scattering angle $\theta$ is the polar angle between the incident electron momentum $\vec{k}_1$ and $\vec{p}_1$.}
    \label{fig:axis}   
\end{figure}

Before proceeding, we introduce the coordinate system employed throughout this work. In the center-of-mass frame of the $e^+$ and $e^-$, the coordinate axes are defined as follows:
\begin{align*}
    \mathbf{e}_x = \frac{1}{\sin \theta} (\hat{k}_1 - \hat{p}_1\cos \theta),  \ \ 
    \mathbf{e}_y = \frac{1}{\sin \theta} \hat{p}_1\times \hat{k}_1, \  \
    \mathbf{e}_z = \hat{p}_1,
\end{align*}
where $\hat{k}_1$ and $\hat{p}_1$ denote the unit vectors along the momenta of the electron and baryon $B$, respectively, and $\theta$ is the scattering angle between them, as shown in Fig.~\ref{fig:axis}.

\subsection{Density matrix of $B\bar{B}$ pair in the reaction $e^+e^-\to B\bar{B}$}

Under the single-photon approximation, the helicity amplitude for $e^+e^-\to B\bar{B}$ reaction is written as
\begin{eqnarray}
    T_{\lambda_1 \lambda_2, \eta_1\eta_2} = \bar{v}(k_2,\eta_2)\gamma_\mu u(k_1,\eta_1)\bar{u}(p_1,\lambda_1)\Gamma^\mu v(p_2,\lambda_2), \label{eq:T}
\end{eqnarray}
where $k_1$, $k_2$ ($q_1$, $q_2$) and $\eta_1$, $\eta_2$ ($\lambda_1$, $\lambda_2$) are the four-momenta and helicities of the electron and positron (baryon $B$ and antibaryon $\bar{B}$), respectively. The interaction vertex $\Gamma^\mu$ for $\gamma^* B\bar{B}$ is parameterized by the Dirac and Pauli form factors, $F_1(q^2)$ and $F_2(q^2)$,
\begin{eqnarray}
\Gamma_\mu = \gamma^\mu F_1(q^2) + i \frac{F_2(q^2)}{2M}\sigma^{\mu\nu}q_\nu ,
\end{eqnarray}
with $q = k_1+k_2$ the four-momentum of the virtual photon $\gamma^*$, and $M$ the mass of baryon $B$. The form factors $F_1$ and $F_2$ are related to the electric and magnetic form factors $G_E$ and $G_M$, via:
\begin{eqnarray}
    G_E &=& F_1(q^2) + \frac{q^2}{4M^2} F_2(q^2), \label{eq:ge} \\
    G_M &=& F_1(q^2) +  F_2(q^2). \label{eq:gm}
\end{eqnarray}

In addition, the helicity eigenstates for a spin-1/2 particle with four-momentum $p = (E, \vec{p})$ and helicity $\lambda = \pm$ are:
\begin{align}
u(\vec{p},\lambda) &= \frac{1}{\sqrt{E+M}}\begin{pmatrix}
        E+M \\
        \lambda |\vec{p}|
    \end{pmatrix} \chi_\lambda(\hat{p}), \nonumber \\
v(\vec{p},\lambda) &= \frac{-\lambda}{\sqrt{E+M}}\begin{pmatrix}
        -\lambda |\vec{p}| \\
        E+M
    \end{pmatrix} \chi_{-\lambda}(\hat{p}),
\end{align}
where $M$ is the particle mass, $\hat{p} = \vec{p}/|\vec{p}| = (\sin\theta \cos\phi, \sin\theta\sin\phi, \cos\theta)$ is the direction of momentum, and $\chi_\lambda(\hat{p})$ are the two-component helicity spinors:
\begin{align}
    \chi_{+}(\hat{p}) = \begin{pmatrix}
        \cos \frac{\theta}{2}e^{-i\frac{\phi}{2}} \\
        \sin \frac{\theta}{2}e^{i\frac{\phi}{2}}
    \end{pmatrix}, \ 
    \chi_{-}(\hat{p}) = \begin{pmatrix}
        -\sin \frac{\theta}{2}e^{-i\frac{\phi}{2}} \\
        \cos \frac{\theta}{2}e^{i\frac{\phi}{2}}
    \end{pmatrix}.
\end{align}

For a particle moving in the $-\hat{p}$ direction, the corresponding spinors are:
\begin{align}
u(-\vec{p},\lambda) &= \frac{1}{\sqrt{E+M}}\begin{pmatrix}
        E+M \\
        \lambda |\vec{p}|
    \end{pmatrix} \chi_{-\lambda}(\hat{p}), \nonumber \\
v(-\vec{p},\lambda) &= \frac{\lambda}{\sqrt{E+M}}\begin{pmatrix}
        -\lambda |\vec{p}| \\
        E+M
    \end{pmatrix} \chi_{\lambda}(\hat{p}).
\end{align}

Taking $\lambda_i,~\eta_i = \pm$ ($i=1,2$), there are 16 helicity amplitudes in total. However, in the center of mass frame, substituting the specific expressions for these spinors into Eq.~\eqref{eq:T}, there are only five independent helicity amplitudes as follows~\footnote{It is clear that neglecting the electron mass ($m_e \approx 0$) leads to $T_{++,++} = T_{-+,++} = 0$.}:
\begin{eqnarray}
    T_{++,++} &=&  T_{++,--} = T_{--,++} = T_{--,--} \nonumber \\
    &=& 4Mm_e \sin \theta G_E, \nonumber \\
    T_{-+,++} &=& T_{-+,--}=-T_{+-,--} = -T_{+-,++} ,  \nonumber \\
    &=&4Em_e \sin \theta G_M,\nonumber \\
    T_{++,+-} &=& -T_{++,-+} = T_{--,+-} = -T_{--,-+} \nonumber \\
    &=&4EM \sin \theta G_E,  \nonumber \\
    T_{+-,+-} &=& T_{-+,-+} = 4E^2(1+\cos \theta)G_M,  \nonumber \\
    T_{+-,-+} &=&  T_{-+,+-} = 4E^2(1-\cos \theta)G_M, 
\end{eqnarray}
where $m_e$ and $E$ are the mass and energy of the electron, respectively. 

For unpolarized initial states $e^-$ and $e^+$, the spin density matrix $\rho_{e^-e^+}$ is $I_4/4$, with $I_4$ the $4 \times 4$ identity matrix. Then, the final $B \bar{B}$ spin density matrix is written as
\begin{eqnarray}
    \rho_{B\bar{B}} = \frac{T\rho_{e^-e^+}T^\dagger}{\mathrm{Tr}[T\rho_{e^-e^+}T^\dagger]}.
\end{eqnarray}
In the chosen reference frame, the produced baryon $B$ moves along the positive $z$-axis, while corresponding $\bar{B}$ recoils in the opposite direction. Thus, $\lambda_1 = +$ ($\lambda_2 = +$) corresponds to spin up (down) for $B$ ($\bar{B}$). The expectation values~\footnote{With $\Theta_{\mu\nu}$, the density matrix $\rho_{B\bar{B}}$ can be rewritten as $\rho_{B\bar{B}} = \frac{1}{4} \Theta_{\mu\nu}\sigma_\mu\otimes \bar{\sigma}_\nu$. Thus, the matrix $\Theta_{\mu\nu}$ is equivalent to $\rho_{B\bar{B}}$ and fully characterizes the spin state of $B\bar{B}$ system.} $\Theta_{\mu\nu}= \braket{\sigma_\mu\otimes \bar{\sigma}_\nu}$  ($\mu$, $\nu$ = 0, $x$, $y$, and $z$) of the spin density matrix can be theoretically calculated with~\cite{Perotti:2018wxm,Zhang:2024rbl,Salone:2022lpt,Cao:2024tvz,Zhang:2025oks,Zhao:2025cbd,Guo:2025bfn,Li:2026bkf}
\begin{eqnarray}
   \Theta_{\mu\nu}= \braket{\sigma_\mu\otimes \bar{\sigma}_\nu} = \mathrm{Tr}[\rho_{B\bar{B}}\sigma_\mu\otimes \bar{\sigma}_\nu]. \label{eq:Theta}
\end{eqnarray}
Here, $\sigma_0 = \bar{\sigma}_0 = I_2$ is the $2 \times 2$ identity matrix, and~\footnote{The convention for $\bar{\sigma}_y$ and $\bar{\sigma}_z$ ensures the correct treatment of the antibaryon's spin in its own rest frame, accounting for its opposite momentum direction relative to the baryon.}
\begin{align}
    \sigma_x &= \bar{\sigma}_x = \begin{pmatrix}
        0 & 1 \\
        1 & 0
    \end{pmatrix},\qquad \nonumber 
    \sigma_y = - \bar{\sigma}_y = \begin{pmatrix}
        0 & -i \\
        i & 0
    \end{pmatrix}, \nonumber \\
    \sigma_z &= - \bar{\sigma}_z = \begin{pmatrix}
        1 & 0 \\
        0 & -1
    \end{pmatrix} . \label{eq:sigmamatrix}
\end{align}

For clarity, one can present $\Theta_{\mu\nu}$ in block-matrix form:
\begin{eqnarray}
    \Theta = \frac{1}{A}\begin{pmatrix}
        A & P_2^T \\ 
        P_1 & C
    \end{pmatrix}. \label{eq:matrixform}
\end{eqnarray}
Here, $A$ is a scalar and it is proportional to the differential cross section, while $P_1$ and $P_2$ are $3\times1$ vectors, and $C$ is a $3\times3$ matrix. Their explicit forms are:
\begin{eqnarray}
    A = 1 + \alpha\cos^2\theta, \qquad P_1 = P_2 = \begin{pmatrix}
        0\\
        \beta \sin \theta \cos \theta \\
        0
    \end{pmatrix}, \\
    C=\begin{pmatrix}
     \sin^2\theta & 0 & \gamma\sin\theta\cos\theta \\
     0 & -\alpha \sin^2\theta & 0\\
     \gamma\sin\theta\cos\theta & 0 & \alpha+\cos^2\theta
\end{pmatrix},
\end{eqnarray}
where $\alpha$ is the angular distribution parameter for $e^+e^-\to B\bar{B}$ reaction, which is
\begin{eqnarray} 
    \alpha = \frac{q^2|G_M|^2 - M^2|G_E|^2}{q^2|G_M|^2 + M^2|G_E|^2}.
\end{eqnarray}
$\beta$ and $\gamma$ are:
\begin{eqnarray}
    \beta = \sqrt{1-\alpha^2}\sin \Delta \Phi, \quad \gamma = \sqrt{1-\alpha^2}\cos \Delta \Phi, \label{eq:betagamma}
\end{eqnarray}
with $\Delta \Phi$ the relative phase between $G_E$ and $G_M$, i.e., $G_EG_M^*=|G_E||G_M|e^{i\Delta \Phi}$. 

The vectors $P_1/A$ and $P_2/A$ are the polarization vectors of $B$ and $\bar{B}$, respectively. The corresponding polarization degrees are
\begin{eqnarray}
    d_1 = \left(\frac{1}{A^2}P_1^TP_1\right)^{\frac{1}{2}}, \qquad d_2 = \left(\frac{1}{A^2}P_2^TP_2\right)^{\frac{1}{2}},
\end{eqnarray}
with $0\leq d_1,d_2\leq1$. Additionally, one can find that the matrix $\Theta$ is symmetric, which stems from the interaction and the $B\bar{B}$ being $CP$ invariant, and the spin is unchanged under the charge conjugation and parity transformation.

Beyond these single-particle polarizations, the produced baryon and antibaryon pair also exhibits spin entanglement. For two-qubit systems, entanglement can be quantified by the entanglement of formation~\cite{Bennett:1996gf}. It increases monotonically with concurrence $\mathcal{C}$~\cite{Hill:1997pfa, Wootters:1997id}, which is a commonly used value to measure entanglement. For example, $ \mathcal{C}=0$ corresponds to a separable (unentangled) state, while $\mathcal{C}=1$ indicates maximal entanglement. In the $e^+ e^- \to B\bar{B}$ reaction, the concurrence for the $B\bar{B}$ system is~\cite{Wu:2024asu}
\begin{eqnarray}
    \mathcal{C} = \frac{ \bigg| 1+\alpha - \sqrt{[1+\alpha \cos (2\theta)]^2 - \beta^2\sin^2 (2\theta)}\bigg|}{2(1+\alpha\cos^2\theta)}. \label{eq:concur}
\end{eqnarray}

\subsection{Density matrix for cascade decay}

If the produced baryon $B$ and antibaryon $\bar{B}$ are unstable, they subsequently decay via $B \to B_1^\prime + P$ and $\bar{B} \to \bar{B}_2^\prime + P$, and their spin density matrices can be extracted from the angular distributions of the decay products~\cite{Wu:2024asu,Perotti:2018wxm,Zhang:2024rbl}, which are given by
\begin{eqnarray}
    t_{\lambda_1^\prime,\lambda_1} &= \bar{u}({q}_1, \lambda_1^{\prime})(a+b\gamma^5)u({p}_1, \lambda_1), \\
    \bar{t}_{\lambda_2^\prime,\lambda_2}&=\bar{v}({p}_2, \lambda_2)(\bar{a}+\bar{b}\gamma^5)v({q}_2, \lambda_2^\prime),
\end{eqnarray}
where $a$, $\bar{a}$ and $b$, $\bar{b}$ are the coupling constants of $BB'P$ and $\bar{B}\bar{B}'P$ vertexes, respectively. And $q_1$, $q_2$ and $\lambda_1^\prime$, $\lambda_2^\prime$ are the four-momenta and helicities of $B_1^\prime$ and $\bar{B}_2^\prime$, respectively. These amplitudes are Lorentz invariant and are evaluated in the rest frames of baryon $B$ and antibaryon $\bar{B}$, which are:
\begin{align}
    t_{+,+} &= A_+ \cos (\frac{\theta_1}{2}) e^{i \frac{\phi_1}{2}}, &  t_{+,-} &= A_+\sin (\frac{\theta_1}{2}) e^{-i\frac{\phi_1}{2}}, \nonumber \\ 
    t_{-,+} &= -A_- \sin (\frac{\theta_1}{2}) e^{i\frac{\phi_1}{2}}, & t_{-,-} &= A_-\cos (\frac{\theta_1}{2}) e^{-i \frac{\phi_1}{2}}. \nonumber \\
    \bar{t}_{+,+} &= -B_+\sin (\frac{\theta_2}{2}) e^{-i\frac{\phi_2}{2}}, & \bar{t}_{+,-} &= -B_+\cos (\frac{\theta_2}{2}) e^{i\frac{\phi_2}{2}}, \nonumber  \\ 
    \bar{t}_{-,+} &= -B_-\cos (\frac{\theta_2}{2}) e^{-i\frac{\phi_2}{2}}, & \bar{t}_{-,-} &= B_-\sin (\frac{\theta_2}{2}) e^{i\frac{\phi_2}{2}}. \nonumber
\end{align}
where the factors $A_\pm$ and $B_\pm$ are given by:
\begin{eqnarray}
    A_+ &= \sqrt{2M}\left( a\sqrt{E_1^\prime+M_1^\prime} -  b\sqrt{E_1^\prime-M_1^\prime} \right), \nonumber\\
    A_- &= \sqrt{2M}\left( a\sqrt{E_1^\prime+M_1^\prime} +  b\sqrt{E_1^\prime-M_1^\prime} \right), \nonumber \\
    B_+ &= \sqrt{2M}\left(\bar{a}\sqrt{E_2^\prime+M_2^\prime} -  \bar{b}\sqrt{E_2^\prime-M_2^\prime} \right), \nonumber \\
    B_- &= \sqrt{2M}\left( \bar{a}\sqrt{E_2^\prime+M_2^\prime} +  \bar{b}\sqrt{E_2^\prime-M_2^\prime} \right). \nonumber
\end{eqnarray}
Here, $M_1^\prime$ and $M_2^\prime$ are the masses of $B_1^\prime$ and $\bar{B}_2$, respectively. The four-momenta $q_1$ in the rest frame of $B$ is given by $q_1=(E_1^\prime, \vec{q}_1)$, with $\vec{q}_1 = |\vec{q}_1|(\sin\theta_1\cos \phi_1, \sin\theta_1\sin\phi_1,\cos\theta_1)$, and 
$q_2$ in the rest frame of $\bar{B}$ is $q_2=(E_2^\prime, \vec{q}_2)$, with $\vec{q}_2 = |\vec{q}_2|(\sin\theta_2\cos \phi_2, \sin\theta_2\sin\phi_2,\cos\theta_2)$. 

Then one can define the decay asymmetry parameters as:
\begin{eqnarray}
    \alpha_1 = \frac{|A_+|^2-|A_-|^2}{|A_+|^2+|A_-|^2},\quad  \alpha_2 = \frac{|B_+|^2-|B_-|^2}{|B_+|^2+|B_-|^2},
\end{eqnarray}
which satisfy $-1\leq \alpha_1, \alpha_2 \leq 1$. The parameters $\alpha_1$ and $\alpha_2$ can be measured from the angular distributions of $B_1^\prime$ and $\bar{B}_2^\prime$, respectively. In addition to $\alpha_1$ and $\alpha_2$, the relative phase $\Delta \Phi_1$ between $A_+$ and $A_-$, and phase $\Delta \Phi_2$ between $B_+$ and $B_-$ are also observable quantities, which are related to the polarization of the $B_1^\prime$ and $\bar{B}_2^\prime$. Analogous to Eq.~(\ref{eq:betagamma}), we can also define:
\begin{eqnarray}
\beta_1 = \sqrt{1-\alpha^2_1}\sin \Delta \Phi_1, & \gamma_1 = \sqrt{1-\alpha^2_1}\cos \Delta \Phi_1, \\
\beta_2 = \sqrt{1-\alpha^2_2}\sin \Delta \Phi_2, & \gamma_2 = \sqrt{1-\alpha^2_2}\cos \Delta \Phi_2.
\end{eqnarray}

Then, from the helicity amplitudes $t_{\lambda_1^\prime,\lambda_1}$ and $\bar{t}_{\lambda_2^\prime,\lambda_2}$, we construct the normalized decay matrices $D_1$ and $D_2$ as:
\begin{align}
    (D_1)_{\lambda_1^\prime,\lambda_1} = \frac{1}{N_1}t_{\lambda_1^\prime,\lambda_1}, \quad (D_2)_{\lambda_2^\prime,\lambda_2} = \frac{1}{N_2}\bar{t}_{\lambda_2^\prime,\lambda_2},
\end{align}
where the normalization factors $N_1$ and $N_2$ are chosen such that:
\begin{align}
    \mathrm{Tr}[D_1D_1^\dagger] = \mathrm{Tr}[D_2D_2^\dagger] = 2.
\end{align}
The spin density matrix for the final decayed $B_1^\prime\bar{B}_2^\prime$ system is then written as:
\begin{eqnarray}
    \rho_{B_1^\prime\bar{B}_2^\prime} = \frac{1}{4W}(D_1\otimes D_2)\rho_{B\bar{B}}(D_1\otimes D_2)^\dagger,
\end{eqnarray}
with the normalization factor
\begin{align*}
    W &= A + \alpha_1P_1^TR_1+\alpha_2P_2^TR_2 + \alpha_1\alpha_2R_1^TCR_2,
\end{align*}
which reflects the angular distribution information of the final $ B_1^\prime$ and $\bar{B}_2^\prime$. And the unit vectors $R_1$ and $R_2$ are:
\begin{align}
    R_1^T = (x_1, y_1, z_1),\qquad R_2^T = (x_2, y_2, z_2),
\end{align}
where $x_i = \sin\theta_i\cos\phi_i$, $y_i = \sin\theta_i\sin\phi_i$, $z_i = \cos\theta_i$ ($i=1,2$), which represent the momentum directions of $B_1^\prime$ and $\bar{B}_2^\prime$ in the rest frames of $B$ and $\bar{B}$, respectively. 

Note that $W$ encapsulates the angular distribution information of the final states $B_1^\prime$ and $\bar{B}_2^\prime$. It explicitly depends on the production parameters $\alpha$ and $\Delta\Phi$, which in turn allows the relative phase $\Delta\Phi$ to be extracted directly from the decay angular distributions. In contrast, $W$ is strictly independent of the decay phases $\Delta\Phi_1$ and $\Delta\Phi_2$, as these purely phase-related parameters influence the spin polarization orientations of the daughter baryons rather than their spatial angular distributions. To reveal this spin information, we express the density matrix in the form:
\begin{align}
    \rho_{B_1^\prime\bar{B}_2^\prime} = \frac{1}{4}\Theta^\prime_{\mu\nu}\sigma_\mu^\prime \otimes \bar{\sigma}_\nu^\prime,
\end{align}
where $\Theta^\prime_{\mu\nu} = \mathrm{Tr}[\rho_{B_1^\prime\bar{B}_2^\prime} \sigma_\mu^\prime \otimes \bar{\sigma}_\nu^\prime]$. In the decays $B \to B_1^\prime+ P$ and $\bar{B} \to \bar{B}_2^\prime +P$, the spin information of $B'$ and $\bar{B}'$ are described by generalized Pauli matrices $\sigma_\mu^\prime$ and $\bar{\sigma}_\nu^\prime$. They are defined in the helicity basis relative to the momentum direction $\hat{q}_i = (\sin\theta_i\cos\phi_i, \sin\theta_i\sin\phi_i, \cos\theta_i)$ of $B_1'$ or $\bar{B}_2'$, respectively:
\begin{align}
    \sigma_x^\prime &= \begin{pmatrix}
        \cos\theta_1\cos\phi_1 & \cos\theta_1\cos\phi_1 - i\sin\phi_1 \\
        \cos\theta_1\cos\phi_1 + i\sin\phi_1 & -\cos\theta_1\cos\phi_1
    \end{pmatrix}, \nonumber \\
    \sigma_y^\prime &= \begin{pmatrix}
        \cos\theta_1\sin\phi_1 & \cos\theta_1\sin\phi_1 + i\cos\phi_1 \\
        \cos\theta_1\sin\phi_1 - i\cos\phi_1 & -\cos\theta_1\sin\phi_1
    \end{pmatrix}, \nonumber \\
    \sigma_z^\prime &= \begin{pmatrix}
        \cos\theta_1 & -\sin\theta_1 \\
        -\sin\theta_1 & -\cos\theta_1
    \end{pmatrix}.
\end{align}
for baryon $B_1^\prime$. And
\begin{align}
    \bar{\sigma}_x^\prime &= \begin{pmatrix}
        \cos\theta_2\cos\phi_2 & \cos\theta_2\cos\phi_2 - i\sin\phi_2 \\
        \cos\theta_2\cos\phi_2 + i\sin\phi_2 & -\cos\theta_2\cos\phi_2
    \end{pmatrix}, \nonumber \\
    \bar{\sigma}_y^\prime &= \begin{pmatrix}
        \cos\theta_2\sin\phi_2 & \cos\theta_2\sin\phi_2 + i\cos\phi_2 \\
        \cos\theta_2\sin\phi_2 - i\cos\phi_2 & -\cos\theta_2\sin\phi_2
    \end{pmatrix}, \nonumber \\
    \bar{\sigma}_z^\prime &= \begin{pmatrix}
        \cos\theta_2 & -\sin\theta_2 \\
        -\sin\theta_2 & -\cos\theta_2
    \end{pmatrix}.
\end{align}
for antibaryon $\bar{B}_2^\prime$. Note that these generalized operators reduce to the standard Pauli matrices when the particles move along the $z$-axis ($\theta_i = 0, \phi_i = 0$).

In a similar manner to Eq.~(~\ref{eq:matrixform}), we express $\Theta^\prime$ as follows:
\begin{align}
    \Theta^\prime = \frac{1}{W}\begin{pmatrix}
        W & {P_2^{\prime}}^T \\
        P_1^\prime & C^\prime
    \end{pmatrix},
\end{align}
where $P_1^\prime/W$ and $P_2^\prime/W$ are the polarization vectors of $B_1^\prime$ and $\bar{B}_2^\prime$, respectively, and $C^\prime$ is the correlation matrix. The components of $\Theta^\prime$ are found to be:
\begin{align}
    P_1^\prime &= \alpha_1R_1(A+\alpha_2P^T_2R_2) + H_1(P_1+\alpha_2CR_2), \label{eq:p1}\\
    P_2^\prime &= \alpha_2R_2(A+\alpha_1P^T_1R_1) + H_2(P_2+\alpha_1CR_1), \label{eq:p2}\\
    C^\prime &= \alpha_1\alpha_2 AR_1R_2^T + \alpha_1 R_1 P^T_2H_2^T + \alpha_2 H_1 P_1 R_2^T \nonumber \\
    &\quad +H_1 C H_2^T, \label{eq:c}
\end{align}
where the matrices $H_1$ and $H_2$ are defined as:
\begin{align}
    H_1 &= \gamma_1 +(1-\gamma_1)R_1R_1^T+\beta_1 K_1, \label{eq:h1} \\
    H_2 &= \gamma_2 +(1-\gamma_2)R_2R_2^T+\beta_2 K_2 ,\label{eq:h2}
\end{align}
with $K_1$ and $K_2$ the antisymmetric matrices:
\begin{align*}
    K_1=
    \begin{pmatrix}
        0 & z_1 & -y_1 \\
        -z_1 & 0 & x_1 \\
        y_1 & -x_1 &0
    \end{pmatrix},\ K_2=
    \begin{pmatrix}
        0 & z_2 & -y_2 \\
        -z_2 & 0 & x_2 \\
        y_2 & -x_2 &0
    \end{pmatrix}.
\end{align*}
The dependence on the decay phases $\Delta\Phi_1$ and $\Delta\Phi_2$ enters through the matrices $H_1$ and $H_2$, via the parameters $\beta_{1,2}$ and $\gamma_{1,2}$. Consequently, these phases influence the correlation matrix $C^\prime$ and also the spin polarization vectors $P_{1,2}^\prime$ of the final states $B_1'$ and $\bar{B}_2'$. 

The spin polarization degrees of the decayed baryon $B'$ and antibaryon $\bar{B}'$ are determined by the magnitudes of their polarization vectors:
\begin{eqnarray}
d_1^\prime = \frac{1}{W} \left| P_1^\prime \right|, \qquad d_2^\prime = \frac{1}{W} \left| P_2^\prime \right|.
\end{eqnarray}
Substituting Eqs.~(\ref{eq:p1}) and (\ref{eq:p2}) yields their explicit forms:
\begin{widetext}
\begin{eqnarray}
d_1^\prime &=& \frac{1}{W} \bigg[ \left( \alpha_1 (A + \alpha_2 P_2^T R_2) + R_1^T (P_1 + \alpha_2 C R_2) \right)^2  + (1-\alpha_1^2) (P_1 + \alpha_2 C R_2)^T (1 - R_1 R_1^T) (P_1 + \alpha_2 C R_2) \bigg]^{1/2}, \label{d1prime} \\
d_2^\prime &=& \frac{1}{W} \bigg[ \left( \alpha_2 (A + \alpha_1 P_1^T R_1) + R_2^T (P_2 + \alpha_1 C R_1) \right)^2  + (1-\alpha_2^2) (P_2 + \alpha_1 C R_1)^T (1 - R_2 R_2^T) (P_2 + \alpha_1 C R_1) \bigg]^{1/2}. \label{d2prime}
\end{eqnarray}
\end{widetext}
From Eqs.~\eqref{d1prime} and \eqref{d2prime}, it is clear that the polarization degrees $d_1^\prime$ and $d_2^\prime$ are independent of the decay phases $\Delta\Phi_1$ and $\Delta\Phi_2$.
For the quantum entanglement, we find that the concurrence of the $B_1^\prime\bar{B}_2^\prime$ system is:
\begin{align}
\mathcal{C}^\prime = \frac{A}{W}\sqrt{(1-\alpha_1^2)(1-\alpha_2^2) }\; \mathcal{C},
\end{align}
where $\mathcal{C}$ is the concurrence of the initial produced $B \bar{B}$ pair in the $e^+ e^-$ annihilation as in Eq.~(\ref{eq:concur}). Besides, since $A$ and $W$ are independent of $\Delta\Phi_1$ and $\Delta\Phi_2$, the final-state concurrence $\mathcal{C}^\prime$ is also independent of these decay phases. For convenience, we define the entanglement evolution factor:
\begin{align}
Z = \frac{\mathcal{C}^\prime}{\mathcal{C}} = \frac{A}{W}\sqrt{(1-\alpha_1^2)(1-\alpha_2^2) }.
\end{align}
The value of $Z$ determines the change in quantum entanglement: $Z > 1$ corresponds to entanglement amplification, while $Z < 1$ corresponds to its reduction.

\subsection{Density Matrix for multi-step cascade decays}

In this section, we consider the scenario in which the produced baryon $B_1^\prime$ and antibaryon $\bar{B}_2^\prime$ are themselves unstable and undergo further decays via $B_1^\prime \to B_1^{\prime\prime}+P$ and $\bar{B}_2^\prime \to \bar{B}_2^{\prime\prime}+P$. Assuming that these subsequent decays follow the same weak decay mode and that $B_1^{\prime\prime}$ ($\bar{B}_2^{\prime\prime}$) carry the same spin and parity as $B_1^\prime$ ($\bar{B}_2^\prime$), we then define the density matrix at the $(n-1)$-th and $n$-th decay steps as:
\begin{align}
    \rho^{(n-1)} &=  \frac{1}{4}\Theta^{(n-1)}_{\mu \nu} \sigma^{(n-1)}_\mu\otimes \sigma^{(n-1)}_\nu, \\
    \rho^{(n)} &=  \frac{1}{4}\Theta^{(n)}_{\mu \nu} \sigma^{(n)}_\mu\otimes \sigma^{(n)}_\nu,
\end{align}
where the $\Theta$ matrices have the block structure:
\begin{align}
\Theta^{(n-1)} &=\frac{1}{W^{(n-1)}}\begin{pmatrix}
W^{(n-1)} & (P_2^{(n-1)})^T\\
P_1^{(n-1)} & C^{(n-1)}
\end{pmatrix}, \\
\Theta^{(n)} &= \frac{1}{W^{(n)}}\begin{pmatrix}
W^{(n)} & (P_2^{(n)})^T\\
P_1^{(n)} & C^{(n)}
\end{pmatrix}.
\end{align}
Here, $W^{(n-1)}$ and $W^{(n)}$ are the angular distribution functions at each stage, proportional to the differential cross sections. The recursive relations connecting $\Theta^{(n)}$ to $\Theta^{(n-1)}$ are:
\begin{widetext}
\begin{align}
W^{(n)} &= W^{(n-1)} + (Q_1^{(n)})^T C^{(n-1)} Q_2^{(n)} + (P_1^{(n-1)})^T Q_1^{(n)}+ (P_2^{(n-1)})^T Q_2^{(n)},  \\
P_1^{(n)} &= Q_1^{(n)}\left(W^{(n-1)}+ (P_2^{(n-1)})^T Q_{2}^{(n)}\right)  + H_1^{(n)}\left(P_1^{(n-1)} + C^{(n-1)} Q_2^{(n)} \right), \label{eq:p1n} \\
P_2^{(n)} &= Q_2^{(n)}\left(W^{(n-1)}+ (P_1^{(n-1)})^T Q_{1}^{(n)}\right)  + H_2^{(n)}\left(P_2^{(n-1)} + (C^{(n-1)})^T Q_1^{(n)} \right), \label{eq:p2n} \\
C^{(n)} &= W^{(n-1)}Q_1^{(n)} (Q_2^{(n)})^T + H_1^{(n)} C^{(n-1)} (H_2^{(n)})^T + Q_1^{(n)} (P_2^{(n-1)})^T (H_2^{(n)})^T + H_1^{(n)} P_1^{(n-1)} (Q_2^{(n)})^T. \label{eq:cn}
\end{align}
\end{widetext}
Here, $Q_1^{(n)} = \alpha_1^{(n)}R_1^{(n)}$ and $Q_2^{(n)} = \alpha_2^{(n)}R_2^{(n)}$, where $\alpha_1^{(n)}$, $\alpha_2^{(n)}$ are the decay asymmetry parameters and $R_1^{(n)}$, $R_2^{(n)}$ are the momentum direction vectors of $B_1^{(n)}$ and $\bar{B}_2^{(n)}$ in the rest frames of $B_1^{(n-1)}$ and $\bar{B}_2^{(n-1)}$, respectively. The matrices $H_1^{(n)}$ and $H_2^{(n)}$ are defined analogously to Eqs.~(\ref{eq:h1}) and (\ref{eq:h2}). The recursion is initialized at $n=0$ with the initial production process of $e^+ e^- \to B \bar{B}$: $W^{(0)} = A$, $P_1^{(0)} = P_1$, $P_2^{(0)} = P_2$, and $C^{(0)} = C$. 

Comparing Eqs.~(\ref{eq:p1}-\ref{eq:c}) with Eqs.~(\ref{eq:p1n}-\ref{eq:cn}), one can see that these equations share identical algebraic structures. The only difference is in the term for $P_2^{(n)}$ (Eq. \ref{eq:p2n}), which uses $(C^{(n-1)})^T Q_1^{(n)}$ instead of $C^{(n-1)} Q_1^{(n)}$. This ensures generality, as the correlation matrix $C^{(n-1)}$ is not guaranteed to be symmetric in multi-step decays, whereas $C$ for the initial production process $e^+ e^- \to B\bar{B}$ is symmetric. 

The recursive relation for the concurrence is given by
\begin{align}
    \mathcal{C}^{(n)} = \frac{W^{(n-1)}}{W^{(n)}} \sqrt{ 1- \left(\alpha_1^{(n)}\right)^2}\sqrt{ 1- \left(\alpha_2^{(n)}\right)^2}\mathcal{C}^{(n-1)}
\end{align}

\section{Spin polarization and entanglement}

This section analyzes the evolution of spin polarization and quantum entanglement throughout the above cascade decays. We summarize six key results that clarify the conditions under which these quantum properties are preserved, enhanced, or reduced.

These six key points are as following:
\begin{enumerate}[(1)]
    \item When $\alpha_1= \alpha_2 =0$ and $\Delta \Phi_1 = \Delta\Phi_2 = 0$, the normalized decay matrices $D_1$ and $D_2$ reduce to unitary operator. Accordingly, the transformation $D_1 \otimes D_2$ acting on $\rho_{B\bar{B}}$ is purely unitary, corresponding to a trivial change of spin basis (a global spatial rotation) that preserves all physical spin observables. We then obtain $P_1^\prime = P_1$, $P_2^\prime = P_2$, and $C^\prime = C$. 
    \item  If $\alpha_1= \alpha_2 =0$, then one can get $d_1^\prime=d_1$, $d_2^\prime=d_2$, and $\mathcal{C}^\prime = \mathcal{C}$. This reveals that the phases $\Delta\Phi_1$ and $\Delta\Phi_2$ cause a reorientation of the polarization vectors, thereby affecting their components but not the polarization degrees $d_{1,2}^\prime$ or the concurrence $\mathcal{C}^\prime$.
    \item If $|\alpha_1|=1$, then $d_1^\prime = 1$; if $|\alpha_2|=1$, then $d_2^\prime = 1$. For both situations, we get $\mathcal{C}^\prime = 0$. This reveals a crucial constraint: maximal parity violation in the decay ($|\alpha_i|=1, \ i =1,2$) forces the corresponding daughter particle into a fully polarized state, independent of the initial polarization configuration. In this extreme case, the spin state of the daughter particle collapses into a pure, separable state with respect to the rest of the system, resulting in the complete loss of quantum entanglement ($\mathcal{C}^\prime=0$). This conclusion is in agreement with recent theoretical results presented in Ref.~\cite{Du:2024sly}.
    \item If $\alpha_2 = 0$, $|\alpha_1|\in (0,1)$, and $d_1\leq \left( 1-\sqrt{1-\alpha^2_1}\right)/|\alpha_1|$, then
    for all $R_1$, one can get $d_{1}^\prime\geq d_1$ and $\mathcal{C}^\prime\leq \mathcal{C}$, which considers an asymmetric case where only the baryon $B$ undergoes a parity-violating decay. The behavior is governed essentially by the initial polarization $d_1$, which is determined by the production angle $\theta$. When $d_1$ is below a certain threshold, the spin polarization of $B_1^\prime$ is always enhanced, whereas its quantum entanglement with $\bar{B}$ is reduced. By contrast, if $d_1$ exceeds this threshold, a subset of particular decay directions $R_1$ exists that lead to reduced the spin polarization and quantum entanglement.
    \item If $|\alpha_1|,|\alpha_2|\in (0,1)$, and $d_1=d_2=0$, then for all $R_1$ and $R_2$, we get $Z \leq 1$. This implies that spin entanglement amplification cannot occur if the initial $B\bar{B}$ pair is unpolarized, since $d_1 = d_2 = \beta\sin\theta\cos\theta/(1+\alpha\cos^2\theta)$, an unpolarized initial state is realized when $\alpha = \pm1$, $\Delta \Phi = 0,\pi$, or $\theta = 0,\pi/2,\pi$. Under any of these conditions, $Z \leq 1$ holds for all decay configurations.
    \item If $|\alpha_1|=|\alpha_2|\in (0,1)$, and $d_1=d_2>0$, then there always exist $R_1$ and $R_2$, such that $Z>1$. This case is particularly noteworthy, as it applies to charge-conjugate decays under $CP$ conservation, for which $\alpha_1 = -\alpha_2$ and hence $|\alpha_1| = |\alpha_2|$. In fact, if the initial state is polarized ($d_1 = d_2 > 0$), one can always find final-state pairs $B_1^\prime\bar{B}_2^\prime$ that the quantum entanglement is amplified ($Z>1$). And the maximum value of $Z$ is obtained as:
\begin{widetext}
    \begin{equation} \label{eq:Zmax}
        Z_{\max} = 
        \begin{dcases}
        \frac{2(1+\alpha\cos^2 \theta)}{{1+\alpha + \sqrt{ (1+\alpha\cos2\theta)^2 - \beta^2\sin^2(2\theta)}} }, &\qquad {\rm for}~~~~ \frac{|\beta\sin2\theta|}{1+\alpha\cos2\theta}< \frac{2|\alpha_B|}{ 1+\alpha_B^2};\\
            \frac{(1+\alpha\cos^2\theta)(1-\alpha_B^2) }{1+\alpha\cos^2\theta-\alpha_B^2\alpha\sin^2\theta - |\alpha_B\beta\sin2\theta|}, &\qquad  {\rm for}~~~~ \frac{|\beta\sin2\theta|}{1+\alpha\cos2\theta}\geq \frac{2|\alpha_B|}{ 1+\alpha_B^2}.
        \end{dcases}
\end{equation}
\end{widetext}
Here, $|\alpha_B| = |\alpha_1|$. The above equation (\ref{eq:Zmax}) confirms that $Z_{\max} > 1$ whenever $|\beta\sin2\theta| > 0$, corresponding to a polarized initial state $B$ with $d>0$. Furthermore, it shows that when $ |\beta\sin2\theta|/(1+\alpha\cos2\theta) < 2|\alpha_B|/(1+\alpha_B^2)$, the maximum quantum entanglement amplification becomes independent of the decay parameter $\alpha_B$.
\end{enumerate}

Physically, the entanglement amplification ($Z>1$) observed in these cascade decays can be interpreted within quantum information theory as a manifestation of \textit{local filtering operations}~\cite{Gisin:1996vmz,Verstraete:2000rnx}. Owing to the intrinsic spin dependence of parity-violating weak decay amplitudes ($\alpha_B \neq 0$), the associated decay matrices $D_1$ and $D_2$ act as non-unitary local filters on the spin states of the parent $B\bar{B}$ pair. Detecting the daughter baryons along specific kinematic directions ($R_1$ and $R_2$) effectively implements a non-unitary post-selection on the initial density matrix. For certain correlated geometric configurations, this selective filtering preferentially suppresses the unentangled, classically mixed components of the initially polarized mixed state. As a result, although the total detection probability $W$ is reduced, the quantum state is distilled, such that the relative weight of entangled components in the surviving sub-ensemble is amplified.

To illustrate these results, we examine the following three processes: 
\begin{enumerate}
    \item $e^+e^-\to J/\psi \to \Lambda(\to p \pi^-)\bar{\Lambda}(\to \bar{p} \pi^+)$ ,
    \item $e^+e^-\to J/\psi \to \Sigma^+(\to p \pi^0)\bar{\Sigma}^-(\to \bar{p} \pi^0)$ ,
    \item $\begin{aligned}[t]
        e^+e^-\to J/\psi &\to \Xi^-(\to \Lambda  \pi^-)\bar{\Xi}^+(\to \bar{\Lambda} \pi^+) \\
        &\to \Lambda (\to p \pi^-) \pi^- \bar{\Lambda}(\to \bar{p}\pi^+)\pi^+.
    \end{aligned}$
\end{enumerate}

Using the available experimental values for these decay parameters and relative phases listed in Table~\ref{tab:parameters}, one can straightforwardly compute the spin polarization and quantum entanglement.

\begin{table}[htbp]
\caption{Parameters for the $e^+e^-\to J/\psi \to \Lambda\bar{\Lambda}$, $\Sigma^+ \bar{\Sigma}^-$, and $\Xi^- \bar{\Xi}^+$ processes and their subsequent decays. The decay asymmetry parameters $\alpha_B$ are given under the assumption of $CP$ conservation.}
\begin{ruledtabular}
\begin{tabular}{cccccc}
        No. &  $B\bar{B}$              & $\alpha$         & $\Delta \Phi$ (rad) & $\alpha_B$          & Refs.\\ \hline
        1.  & $\Lambda \bar{\Lambda}$   & 0.461$\pm$0.09   & 0.740$\pm$ 0.014    & 0.754 $\pm$ 0.011   &\cite{BESIII:2018cnd}\\
        2.  & $\Sigma^+ \bar{\Sigma}^-$ & -0.508$\pm$ 0.07 & -0.270$\pm$ 0.015   & -0.994 $\pm$ 0.038  &\cite{BESIII:2020fqg}\\
        3.  & $\Xi^- \bar{\Xi}^+$       & 0.586$\pm$ 0.016 & 1.213$\pm$ 0.048    & -0.374 $\pm$ 0.008  &\cite{BESIII:2021ypr}\\
\end{tabular}
\label{tab:parameters}
\end{ruledtabular}
\end{table}

\begin{figure}[htbp]    
    \centering
    \includegraphics[scale=0.38]{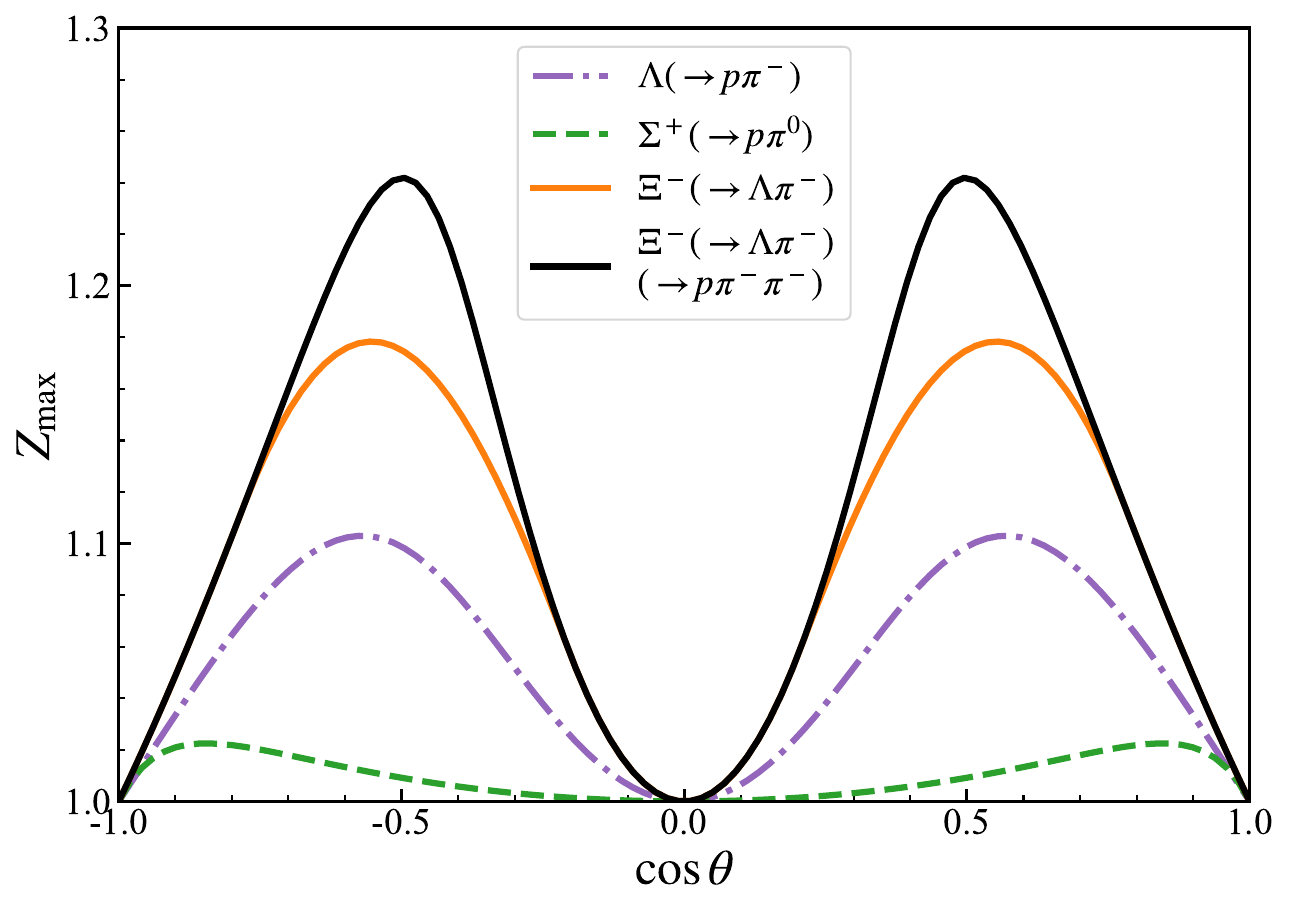}
    \caption{Maximum entanglement amplification factor $Z_{\max}$ as a function of $\cos\theta$ for the final proton in the $J/\psi \to \Lambda\bar{\Lambda}$, $\Sigma^+ \bar{\Sigma}^-$, and $\Xi^- \bar{\Xi}^+$ processes with subsequent weak decays. $\Lambda\bar{\Lambda}$ (purple dash-dotted) with $\Lambda\to p\pi^-$, $\Sigma^+\bar{\Sigma}^-$ (green dashed) with $\Sigma^+\to p\pi^0$, $\Xi^-\bar{\Xi}^+$ (orange solid) with $\Xi^-\to\Lambda\pi^-$, and the multi-step cascade decays of the $\Xi^-\bar{\Xi}^+$ (black solid) with $\Xi^-\to \Lambda \pi^- \to p \pi^-\pi^-$. The corresponding charge-conjugate decays are implied. }
    \label{fig:Zmax}   
\end{figure}

Numerical results for the maximum quantum entanglement amplification factor $Z_{\max}$ of the final proton, as a function of $\cos\theta$, are presented in Fig.~\ref{fig:Zmax} for these three processes discussed above. For $|\cos \theta| \in (0,1)$, where the initial state is polarized ($d>0$), the $Z_{\max}$ exceeds 1, demonstrating that quantum entanglement amplification is achievable. At $\theta= 0, \pi/2, \pi$, where initial polarization vanishes, $Z_{\max} = 1$. The $\Xi^-$ process exhibits the strongest amplification, primarily due to its larger $|\sin(\Delta\Phi)|$ value relative to the $\Lambda$ and $\Sigma^+$ processes. These results imply that the quantum entanglement amplification in baryon-antibaryon pairs produced in $e^+e^-$ annihilation is more pronounced for initially more strongly polarized states.

Furthermore, if we consider the multi-step cascade scenario where $\Xi^-$ decays into $\Lambda \pi^-$, and $\Lambda$ subsequently decays into $p\pi^-$, the maximum enhancement factor $Z_{\rm max}$ for the entanglement of the final $p\bar{p}$ pair relative to the initial $\Xi^-\bar{\Xi}^+$ pair is illustrated by the black line. It is evident that the entanglement amplification still occurs ($Z_{\rm max} > 1$). Comparing the black line with the orange line (which represents the single-step decay), the value of $Z_{\rm max}$ is larger for the multi-step process around $|\cos \theta|\approx 0.5$, indicating that entanglement can be further amplified through sequential cascade decays. In contrast, in the kinematic regions near $ \cos\theta = 0,\pm 1$, the two curves coincide. This indicates that multi-step cascade decays can continue to amplify entanglement in certain kinematic regions, whereas in others, the amplification saturates at the upper limit reached after the primary decay. The underlying reason is that, although the successive decays share similar weak decay mechanisms, the intermediate spin density matrix of the daughter baryon pair differs fundamentally from that of the initially produced $B\bar{B}$ pair. Consequently, the distinct pre-decay spin states directly dictate the subsequent evolution of entanglement, explaining why further amplification is possible in specific angular regions while being constrained by the primary decay's upper limit elsewhere.

\section{Summary}

In this work, we have conducted a systematic investigation into the evolution of the spin polarization and quantum entanglement through sequential weak decays in the process $e^+e^- \to B\bar{B} \to (B \to B_1^\prime+ P)(\bar{B} \to \bar{B}_2^\prime + P)$. A central result of this study is the derivation of a fully analytical spin density matrix for the final-state baryon-antibaryon pair ($B_1^\prime\bar{B}_2^\prime$), expressed explicitly in terms of experimentally accessible spin polarization observables. We further generalize this framework by deriving compact recursive relations that enable straightforward computation of the density matrix for multi-step cascade decays.

Our analysis clarifies how decay parameters govern the transfer and transformation of spin information. It is found that in decays with maximal parity-violation ($|\alpha|=1$), the daughter baryon is produced in a fully polarized, separable state with no residual quantum entanglement. Moreover, we establish explicit conditions under which polarization and entanglement are either amplified or suppressed following the decay. A key result is that entanglement amplification ($Z>1$) is strictly forbidden if the initial $B\bar{B}$ pair is unpolarized. By contrast, for polarized initial states—a generic outcome in $e^+e^-$ annihilation—entanglement amplification always arises in charge-conjugate decays under $CP$ conservation. This effect is verified both analytically and via numerical examples, including $J/\psi \to \Xi^-\bar{\Xi}^+$ and $\Lambda\bar{\Lambda}$, in which substantial entanglement amplification is observed.

\section{ACKNOWLEDGMENTS}

We would like to thank Prof. Rong-Gang Ping for useful discussions. This work is partly supported by the National Key R\&D Program of China under Grant No. 2023YFA1606703, and by the National Natural Science Foundation of China under Grant Nos. 12575094, 12435007 and 12361141819.

\setcounter{equation}{0}
\renewcommand{\theequation}{A\arabic{equation}}
\setcounter{figure}{0}
\renewcommand{\thefigure}{A\arabic{figure}}

\bibliographystyle{apsrev4-1}
\bibliography{ref}
\onecolumngrid

\end{document}